\documentstyle[epsf,sprocl]{article}

\def\cpc#1#2#3 {{\em Computer Phys.\ Comm.}\ {\bf #1}, #2 (19#3)}

\def\np#1#2#3{{\em Nucl.\ Phys.}\ B {\bf #1}, #2 (19#3)}
\def\npps#1#2#3{{\em Nucl.\ Phys. Proc. Suppl. }\ {\bf #1}, #2 (19#3)}
\def\pl#1#2#3{{\em Phys.\ Lett.}\ B {\bf #1}, #2 (19#3)}
\def\pr#1#2#3{{\em Phys.\ Rev.}\ D {\bf #1}, #2 (19#3)}

\def\prl#1#2#3{{\em Phys.\ Rev.\ Lett.}\ {\bf #1}, #2 (19#3)}

\def\zp#1#2#3{{\em Z.\ Phys.}\ C {\bf #1}, #2 (19#3)}

\def\beq{\begin{equation}}
\def\eeq{\end{equation}}
\def\bea{\begin{eqnarray}}
\def\eea{\end{eqnarray}}

\def\mafigura#1#2#3#4{
  \begin{figure}[hbtp]
    \begin{center}
      \epsfxsize=#1
      \leavevmode
      \epsffile{#2}
    \end{center}
    \caption{#3}
    \label{#4}
\end{figure} }

\newcommand{\eq}[1]{eq.~(\ref{#1})}
\newcommand{\as}{\alpha_s}

\newcommand{\yc}{y_{c}}
\newcommand{\rb}{r_b}

\begin{document}

\title{DO QUARK MASSES RUN?}

\author{Arcadi SANTAMARIA}

\address{Departament de F\'{\i}sica Te\`orica, IFIC,
Universitat de Val\`encia-CSIC, 46100 Burjassot, Val\`encia, Spain}

\author{Germ\'an RODRIGO\footnote{
Supported in part by CSIC-Fundaci\'o Bancaixa.
On leave from Departament de F\'{\i}sica Te\`orica, IFIC,
Universitat de Val\`encia-CSIC, 46100 Burjassot, Val\`encia, Spain.}
}

\address{Institut f\"ur Teoretische Teilchenphysik, Universit\"at
Karlsruhe, 76131 Karlsruhe, Germany}

\author{Mikhail BILENKY\footnote{On leave from JINR, 141980 
Dubna, Russian Federation.}
}

\address{Institute of Physics, AS CR, 18040 Prague 8, and 
Nuclear Physics Institute, AS CR, 25068 \v{R}e\v{z}(Prague),
Czech Republic}


\maketitle\abstracts{The importance
of measuring the b-quark mass at different scales is emphasized.
The recent next-to-leading order calculation of three jet heavy quark 
production at LEP and its use to measure $m_b(m_Z)$ is discussed.}

\section{Introduction}

The question of the origin of the masses of quarks and leptons
is one of the unresolved
puzzles in present high energy physics. To answer this question one needs
to know precisely their value. However, quarks are not free and their mass
has to be interpreted more like a coupling than an inertial parameter
and it can run if measured at
different scales. Moreover, in the standard model (SM) all
fermion masses come from Yukawa couplings and those also run with
the energy. To test fermion mass models one has to run masses
extracted at quite different scales to the same scale and compare them
with the same ``ruler''. This way, for instance, one can check that in some
unified models the $b$-quark mass and the $\tau$-lepton mass, although
different at threshold energies they could be equal at the unification scale.
For instance, in fig.~\ref{fig:running} we plot the evolution of the
b-quark and the $\tau$-lepton masses for both the standard model and
the mininal supersymmetric standard model (MSSM). We see that the MSSM
behaves much better than the SM since ``unification'' of
masses could happen at much higher energies in the MSSM.
\mafigura{11.5cm}{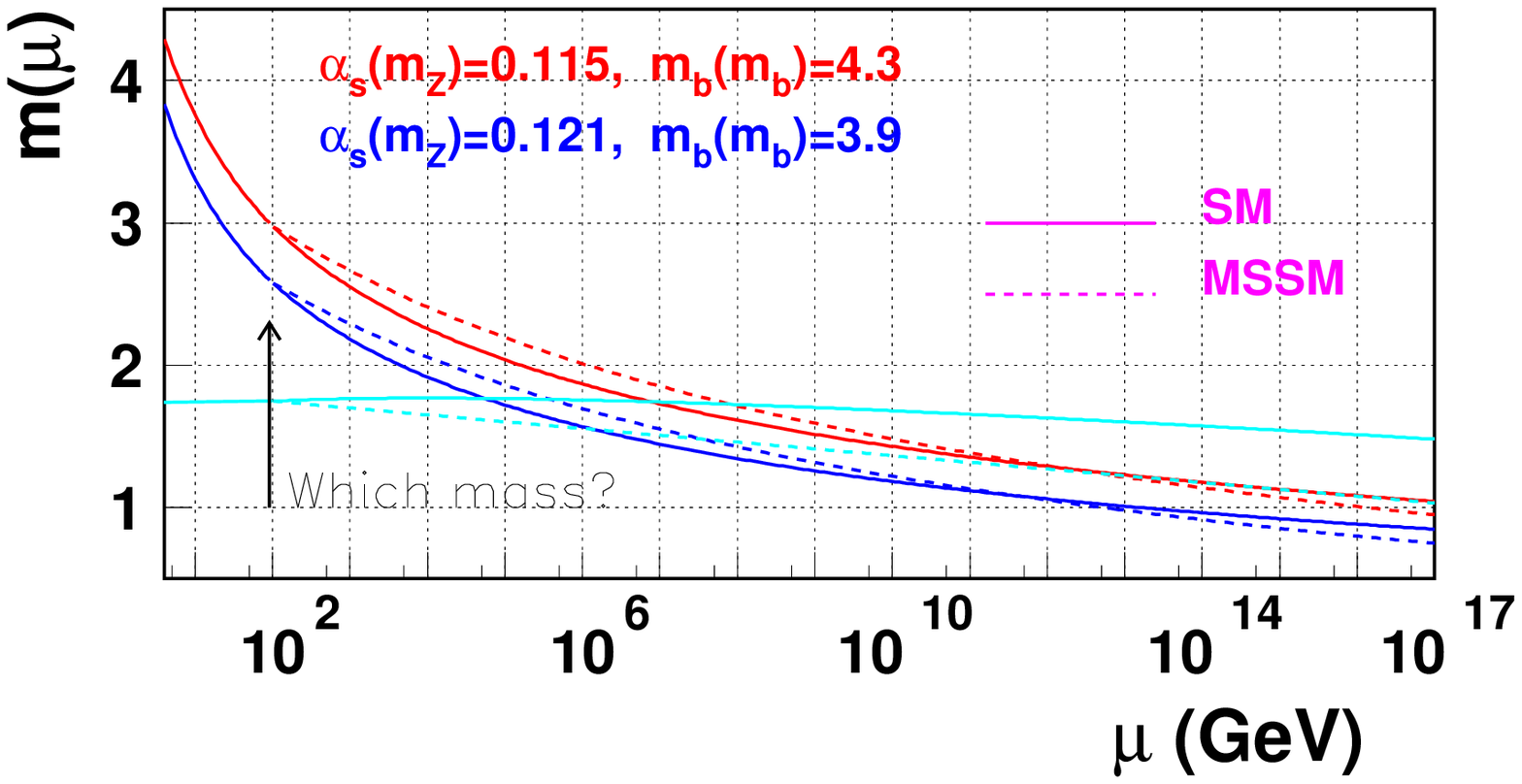}{
Running of the b-quark (dark) and $\tau$-lepton (gray) masses for the standard model
(solid) and the MSSM (dashed). b-quark mass running if given for two values of
the strong couling constant and the mass.}{fig:running}

An important part of the running occurs at energies accessible in 
present experiments. 
However, the running of fermion masses, although predicted by
quantum field theory, has not been tested experimentally until now.
The reason being that for energies $\sqrt{Q^2}$ much higher than
the fermion mass, $m_q$, the mass effects become negligible since
usually they are suppressed by $m_q^2/Q^2$. 
But this argument is not always correct. In fact, in the SM,
the partial decay width of the Higgs-boson into b-quarks is proportional
to $m_b^2(m_H)$ because the Yukawa coupling of the Higgs-boson.
In addition, it is not true~\cite{ballestrero}
for jet cross sections which depend
on a new variable, $\yc$ (the jet-resolution parameter
that defines the jet multiplicity) and
which introduces a new scale in the analysis, $E_c=\sqrt{Q^2\: \yc}$. Then,
for small values of $\yc$ there could be contributions coming like
$m_q^2/E_c^2 = (m_q^2/Q^2) /\yc$ which could enhance the mass effect
considerably. In addition mass effects could also be enhanced by
logarithms of the mass. For instance, the ratio of the phase space
for two massive quarks and a gluon to the phase
space for three massless
particles is $1+8 (m_q^2/Q^2) \log(m_q/Q)$.
At $Q^2=m_Z^2$ and for the bottom quark this gives a 7\% effect,  
for $m_b=5$~GeV and a 3\% effect for $m_b=3$~GeV.
The high precision achieved at LEP makes these effects relevant.
We showed~\citelow{bilenky.rodrigo.ea:95} that b-quark mass effects
in three-jet production at LEP are large enough
to be measured and, in fact, they were already observed 
in the tests of the flavour independence of $\as(m_Z)$~\cite{chrin,l3}.
Therefore, these effects could be used to measure the b-quark mass at LEP.

The observable proposed as a means to extract
the bottom-quark mass from LEP data was the 
ratio~\cite{bilenky.rodrigo.ea:95,chrin}
\begin{equation}
R^{bd}_3 \equiv \frac{\Gamma^b_{3j}(\yc)/\Gamma^b}
{\Gamma^d_{3j}(\yc)/\Gamma^d}~.
\label{eq:r3bd_def}
\end{equation}
In this equation $\Gamma^q_{3j}(\yc)/\Gamma^q$ is the three-jet fraction
of $Z$-decays and $q$ denotes the quark flavor.
Obviously, the ratio $R^{bd}_3$ depends on the jet-clustering 
algorithm~\cite{bethke.kunszt.ea:92} used to define the jets.
In this ratio and at the leading order (LO) the quark mass effects 
can be as large as 1\% to 6\%,
depending on the values of the mass and 
the jet-resolution parameter, $\yc$.

Since the measurement of $R^{bd}_3$ is done
far away from the threshold of $b$-quark production, it can, in principle,
be used to test the running of a quark mass as predicted by QCD.
However,
the leading order calculation does not distinguish among the
different definitions of the quark mass~\citelow{bilenky.rodrigo.ea:95}: 
perturbative pole mass, $M_b$,
running mass at $M_b$-scale, or running mass at $m_Z$-scale. 
Therefore, to distinguish them it is necessary to 
use a complete next--to--leading order (NLO) calculation of 
three-jet ratios including quark masses which was not available until
very recently~\cite{rodrigo:96,rodrigo.santamaria.ea:97*a,rodrigo.santamaria.ea:97*b,bernreuther.brandenburg.ea:97,nason.oleari:97}.
Here we overview the calculation~\cite{rodrigo.santamaria.ea:97*a}
used by the DELPHI Collaboration to extract the 
$b$-quark mass at the $m_Z$ scale~\cite{fuster.cabrera.ea:96,marti.fuster.ea:97,delphinou}.

\section{Jet ratios with heavy quarks at NLO}

The decay width of the $Z$-boson into three jets with a heavy 
quark can be written as follows 
\beq
\Gamma^{b}_{3j} = \frac{m_Z g^2 \alpha_s}{c_W^2 64 \pi^2}
\left[g_V^2 H_V(y_c,r_b) + g_A^2 H_A(y_c,r_b)\right]~,
\label{eq:gamma3jets}
\eeq
where $g$ is the SU(2) gauge coupling constant,
$c_W$ and $s_W$
are the cosine and the sine of the weak mixing angle,
$g_V =-1+4/3 s_W^2$ and $g_A=1$ are the vector and axial-vector
coupling of the
$Z$-boson to the bottom quark
and $\alpha_s$ is the strong coupling constant.
Functions $H_{V(A)}(\yc,\rb)$
contain all the dependences on $\yc$ and the quark
mass, $\rb = (M_b/m_Z)^2$, for the different
algorithms. These functions are computed perturbatively as an expansion
in $\as$ and can also be expanded in $\rb$ for $\rb << 1$.
At the NLO we have contributions to the three-jet cross section from three-
and four-parton final states.
One-loop three-parton amplitudes are both infrared (IR) and ultraviolet (UV)
divergent. However, the UV divergences are removed after renormalization.
The four-parton transition amplitudes are also IR divergent but the sum
of both, four-parton and three-parton, contributions is IR finite. We use the 
so-called {\it phase space slicing method}~\cite{slicing} to obtain the 
remaining finite results, that is, the functions
$H_{V(A)}$ in \eq{eq:gamma3jets} at order $\as$. These results have been
checked in different ways, in particular we checked that the 
massless result~\cite{bethke.kunszt.ea:92,ert}
is nicely recovered when we take the limit $M_b \rightarrow 0$. These results
have also been checked independently by two different
groups~\cite{bernreuther.brandenburg.ea:97,nason.oleari:97}.

Combining
\eq{eq:r3bd_def}, 
\eq{eq:gamma3jets} 
and using the
known expression for $\Gamma^b$ 
\cite{bilenky.rodrigo.ea:95,djouadi.kuhn.ea:90}
we write $R^{bd}_3$ as the following expansion in $\as$ 
\begin{equation}
R^{bd}_3 = 1+ \rb \left(b_0+\frac{\as}{\pi} b_1\right)~,
\label{eq:r3resultpole}
\end{equation}
where the functions $b_0$ and $b_1$ are given by an average of the
vector and axial-vector parts of the Z-widths, weighted by
$c_V= g_V^2/(g_V^2+g_A^2)$ and
$c_A= g_A^2/(g_V^2+g_A^2)$ respectively. 
It is important to note that, because of the particular normalization we have
used in the definition of $R^{bd}_3$, most of the electroweak corrections 
cancel. 

Although intermediate calculations have been performed using the 
pole mass, we can also re-express our results in terms of the running
quark mass by using the known perturbative expression
$
M_b^2 = m_b^2(\mu) [1 + 2\as(\mu)/\pi\ 
( 4/3 - \log(m_b^2/\mu^2) ) ]~.
$
We obtain
\begin{equation}
R^{bd}_3 = 1+ \bar{r}_b(\mu) \left(
b_0+\frac{\as(\mu)}{\pi} 
\left[\bar{b}_1-2 b_0 \log \frac{m_Z^2}{\mu^2}\right]
\right)~,
\label{eq:r3resultrunning}
\end{equation}
where $\bar{r}_b(\mu)=m^2_b(\mu)/m_Z^2$ and
\begin{equation}
\bar{b}_1 = b_1+b_0\left[8/3-2\log(\rb)\right]~.
\label{eq:b1bar}
\end{equation}
$\bar{r}_b(\mu)$ can be expressed in terms of the running mass of
the $b$-quark at $\mu = m_Z$ by using the renormalization group. At the
order we are working
$
\bar{r}_b(\mu)=\bar{r}_b(m_Z)\left(
\alpha_s(m_Z)/\alpha_s(\mu)\right)^{-4\gamma_0/\beta_0}
$
with
$
\alpha_s(\mu) = \alpha_s(m_Z)/(1+\alpha_s(m_Z)\beta_0 t) 
\label{eq:mbrunning}
$
and $t=\log(\mu^2/m^2_Z)/(4\pi)$, $\beta_0 =11-2N_f/3$, $N_f=5$ 
and $\gamma_0=2$.
\mafigura{7cm}{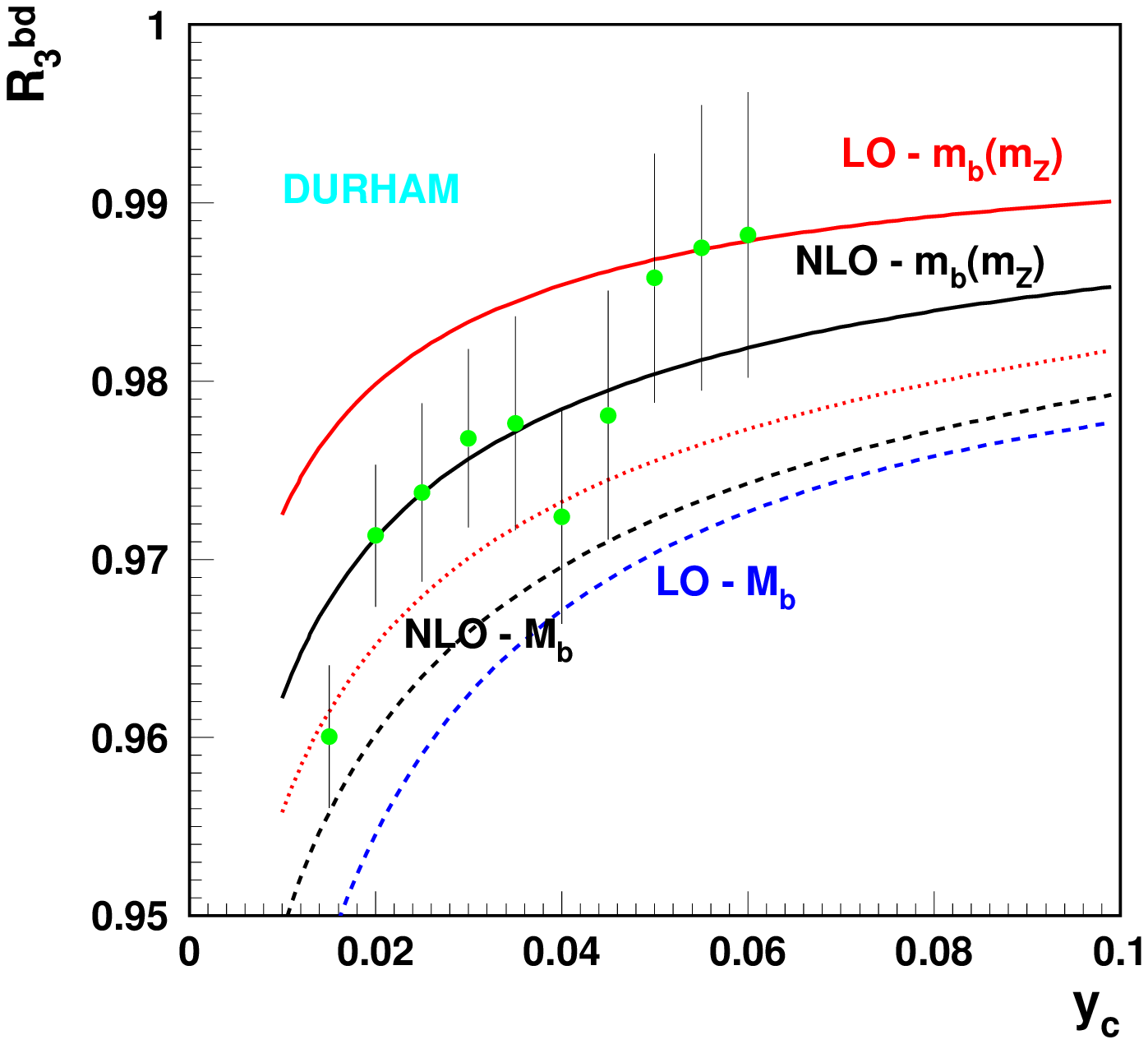}{
NLO results for $R_3^{bd}$ (DURHAM) written in terms of the pole
mass, $M_b$ (dashed), or the running mass $m_b(m_Z)$ (solid) at
LO (gray) and at the NLO (dark). The dotted line gives the NLO result
in terms of the running mass at an intermediate scale $m_b(10~GeV)$. We use 
as a starting point $m_b(m_b)=4.13$ to obtain both $M_b$ and $m_b(m_Z)$.}
{fig:r3}

At the perturbative level \eq{eq:r3resultpole} and \eq{eq:r3resultrunning}
are equivalent. However, they neglect different higher order terms
and lead to different answers. The difference gives an estimate of the
size of higher order corrections.

Here we present only the results for the DURHAM algorithm 
\cite{brown.stirling:92*b},
which gives smaller radiative
corrections and was the one used by the
DELPHI Collaboration in its analysis.

The function $b_0$ describes the corrections due to the quark mass at LO and 
has an almost negligible residual mass dependence. 
A fit, $b_0 = \sum_{n=0}^2 k_0^{(n)}$ log${}^n y_c$, 
to the complete results for the DURHAM algorithm gives:
$k_0^{(0)}=-10.521\;$,  $k_0^{(1)}=-4.4352 \;$, $k_0^{(2)}=-1.6629\;$.

The function~\cite{rodrigo.santamaria.ea:97*a} $\bar{b}_1$
gives the NLO massive corrections to $R^{bd}_3$. 
It is important to note that $\bar{b}_1$
contains significant logarithmic corrections depending on the quark mass. 
We take them into account by using the form
$\bar{b}_1= k_1^{(0)} + k_1^{(1)}\log(\yc) + k_m^{(0)} \log (\rb)$ 
in the fit. For the DURHAM scheme we obtain:
$k_1^{(0)}=297.92\;$,  $k_1^{(1)}=59.358 \;$, $k_m^{(0)}=46.238\;$.

In fig.~\ref{fig:r3} we present $R_3^{bd}$ as a function of $y_c$
for DURHAM. To compute it, we use the low energy 
measurement of the b-quark mass~\cite{jamin.pich:97}, $m_b(m_b)=4.13$. 
Note that choosing a low value for $\mu$ in the NLO predictions written in 
terms of the running mass makes it closer to the LO result written
in terms of the pole mass, while choosing a large $\mu$ makes the result
approach to the LO result written in terms of the running mass at the
$m_Z$ scale.

\section{$m_b(m_Z)$ from LEP data}

Since $R_3^{bd}$ has been measured to good accuracy by DELPHI~\cite{delphinou}
one can use \eq{eq:r3resultrunning} and the relationship between
$m_b(\mu)$ and $m_b(m_Z)$
to extract $m_b(m_Z)$. On the other hand, one could use
\eq{eq:r3resultpole} to obtain directly the pole mass, $M_b$, and then use
the relationship between the pole mass and the running mass and the
renormalization group to obtain also $m_b(m_Z)$. Both results
are slightly different and have a residual dependence on the scale $\mu$.
The difference gives an estimate of the errors due to the unknowledge of
higher order corrections. In fig.~\ref{fig:mbbar} we plot, as a function of
$\mu$, the values of  $m_b(m_Z)$ obtained from $R^{bd}_{3\, exp}=0.971$,
by using the two methods. The most conservative
estimate of the theoretical error is obtained by 
taking the spread of the results at $\mu=m_Z$.
\mafigura{7cm}{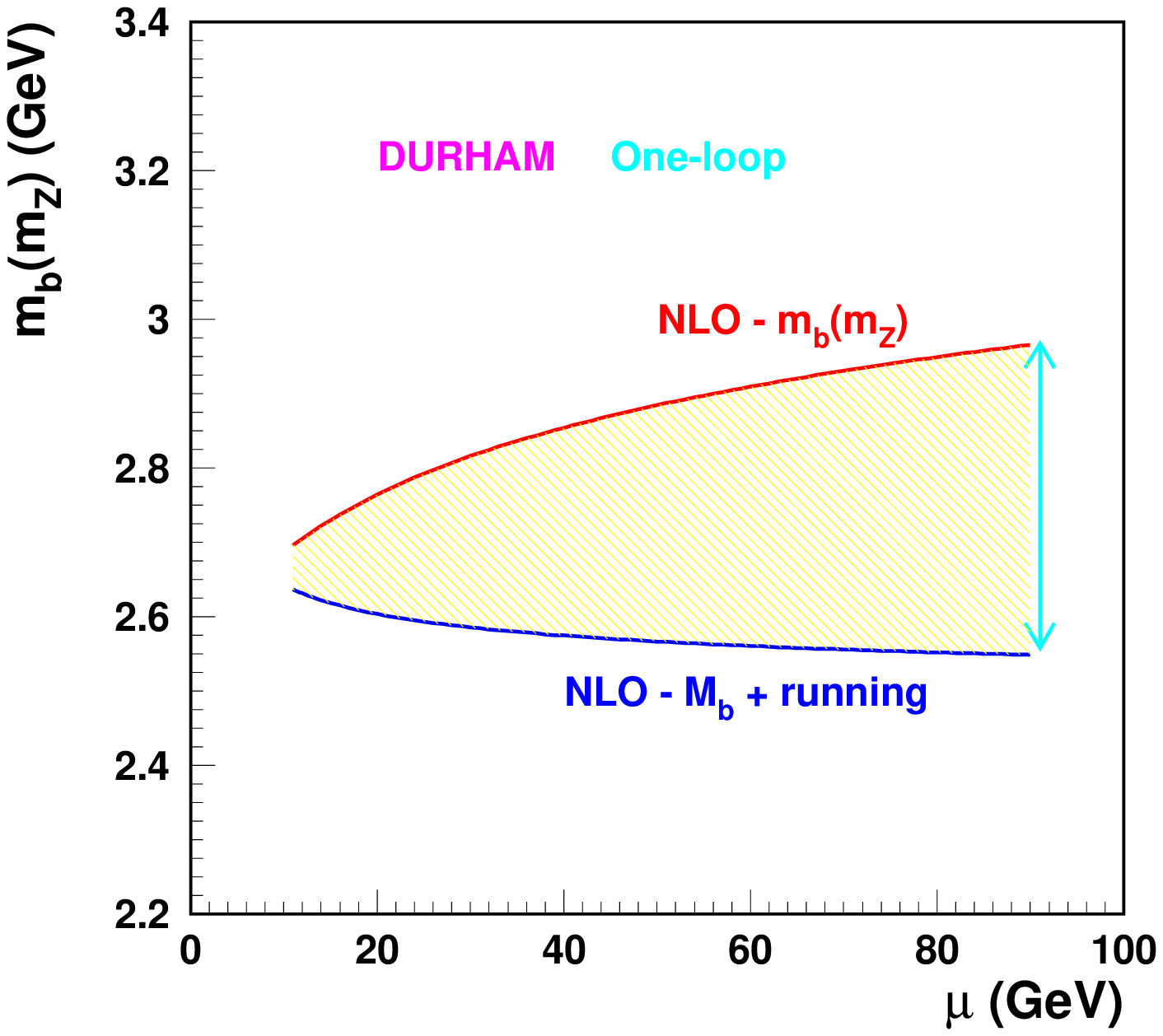}{
Extracted value of $m_b(m_Z)$ if $R^{bd}_{3\, exp}=0.971$ as
a function of the scale $\mu$ by using both the pole mass result
and the running mass result.}{fig:mbbar}

The final result, after including also statistical errors and the
errors due to the uncertainties in the hadronization corrections has been
presented by DELPHI in this conference~\cite{delphinou}.
The obtained mass of the bottom-quark, $m_b(m_Z)$, measured from the
three-jet decay of the Z-boson~\cite{delphinou}
is fully compatible with the value
obtained from low energy determinations~\cite{jamin.pich:97}
after using the renormalization group.
This provides, for the first time,
a nice check of the quark mass sector of QCD in a very wide range of
scales. These results can probably be improved by understanding better
the scale dependence of the results, by resummation of large logs,
by using other observables with a softer dependence on the scale,
by reducing the hadronization uncertainties and finally by including all
LEP data. We think that these studies can provide a rather precise value
of $m_b(m_Z)$.

\end{document}